\documentstyle[aps,preprint]{revtex}
\tightenlines
\begin{document}

November 1997 \hspace{3.0 in} UAS-FCFM/97-6

\medskip\ 

\bigskip\ 

\begin{center}
{\hyphenation{des-cu-bri-mien-to}
\hyphenation{Mo-der-na}
\hyphenation{na-tu-ra-le-za}
\hyphenation{cua-tro}
\hyphenation{co-ti-dia-na}
\hyphenation{In-te-raccio-nes}
\hyphenation{nues-tro}
\hyphenation{di-fe-ren-tes}
\hyphenation{dua-li-dad}
\hyphenation{atri-bu-ye-ron}
\hyphenation{me-so-ame-ri-ca-nos}
}

{\bf \ MATROID THEORY AND SUPERGRAVITY}

\bigskip\ 

J. A. Nieto\footnote{%
E-mail address: nieto@uas.uasnet.mx}

\medskip\ 

{\it Facultad de Ciencias F\'{\i}sico-Matem\'{a}ticas }

{\it de la Universidad Aut\'{o}noma de Sinaloa, }

{\it C.P. 80010, Culiac\'{a}n, Sinaloa, M\'{e}xico.}

\medskip\ 

\date{\today}
\bigskip\ 

{\bf Abstract}
\end{center}

In this work, we consider matroid theory. After presenting three different
(but equivalent) definitions of matroids, we mention some of the most
important theorems of such theory. In particular, we note that every matroid
has a dual matroid and that a matroid is regular if and only if it is binary
and includes no Fano matroid or its dual. We show a connection between this
last theorem and octonions which at the same time, as it is known, are
related to the Englert's solution of D = 11 supergravity. Specifically, we
find a relation between the dual of Fano matroid and D = 11 supergravity.
Possible applications to M-theory are speculated upon.


Pacs No.: 04.50.+h and 04.90.+z

\newpage\ 

At present, the concept of duality is widely recognized by its central role
in non-perturbative dynamics of superstrings [1] and supersymmetric
Yang-Mills [2]. In particular, the five known superstring theories (called
Type I, Type IIA, Type IIB, Heterotic SO(32) and Heterotic E$_8\times $ E$_8$%
) may now be thought, thanks to duality, as different vacua of an underlying
unique theory called M-theory [3]-[9]. This feature of duality in
superstring theories is so relevant that lead us to believe that there must
be a duality principle supporting M-theory.

M-theory is a non-pertubative theory and in addition to the five superstring
theories describes supermembranes [10], 5-branes [11] and D = 11
supergravity [12]. Although the complete M-theory is unknown there are two
main proposed routes to construct it. One is the N = (2,1) superstring
theory [13] and the other is M(atrix)-theory [14]. Recently, Martinec [15]
has suggested that these two scenarios may in fact be closely related.

In this work we propose an alternative formalism to construct M-theory. We
propose that the mathematical formalism necessary to support the duality
principle in M-theory is matroid theory [16]. As it is known, matroid theory
can be understood as a generalization of matrix and graph theory and among
its remarkable features is that every matroid has its dual. Since
M(atrix)-theory and N = (2,1) superstrings have had an important success in
describing some essential aspects of M-theory a natural question is to see
whether matroid theory is related to these two approaches. As a first step
to answer this question we may attempt to investigate if matroid theory is
connected somehow to D = 11 supergravity, which is a common feature of both
approaches. In this work, we find that the Fano matroid and its dual are
related to Englert's compactification [17] of D = 11 supergravity. This
relation is physically interesting for at least two reasons. First, since in
matroid theory the Fano matroid plays a fundamental role [18] we should
expect that matroids may be helpful to describe some important properties of
D = 11 supergravity. Second, it turns out that such a relation allows to
connect the Fano matroid with octonions (one of the division algebras [19])
which are possible related with the four forces of nature. In fact, some
time ago, Blencowe and Duff [20] raised the question whether the four forces
of nature correspond to the four division algebras. If this conjecture turns
out to be true then, according to our results, matroids must be deeply
connected with the four forces of nature.

Let us start with a brief historical review of matroid theory. It seems that
the theory began in 1935 with Whitney's paper [16]. In the same year,
Birkhoff [21] established the connection between simple matroids (also known
as combinatorial geometries [22]) and geometric lattices. In 1936, Mac Lane
[23] gave an interpretation of matroids in terms of projective geometry. And
important progress was made in 1958, with two Tutte's papers [18]. At
present, there is a large body of information about matroid theory and the
closely related combinatorial geometries. Concrete applications may be found
in circuit theory, network-flow theory, linear and integer programming and
the theory (01)-matrices, for example. For further details about the history
of matroid theory and related topics see, for example, the excellent books
by Welsh [24], Lawler [25] and Tutte [26]; and also by Wilson [27], Kung
[28] and Ribnikov [29].

An interesting feature of matroid theory is that there are many different
but equivalent ways of defining a matroid. In this respect, it turns out
interesting to briefly review Whitney's [16] original discovery of the
matroid concept. While working with linear graphs, Whitney noticed that, for
certain matrices, duality had a simple geometrical interpretation quite
different that in the case of graphs. He also observed that any subset of
columns of a matrix is either linearly independent or linearly dependent and
that the following two statements must hold:

(a) any subset of an independent set is independent.

(b) if N$_p$ and N$_{p+1}$ are independent sets of p and p+1 columns
respectively, then N$_p$ together with some column of N$_{p+1}$ forms an
independent set of p+1 columns.

\noindent Moreover, Whitney discovered that if these two statements are
taken as axioms then there are examples of systems satisfying these axioms
but not representing any matrix or graph. Thus, he concluded that a system
obeying (a) and (b) should be a new one and therefore deserved a new name:
matroid.

The definition of a matroid (o pregeometry) in terms of independent sets has
been refined and it is now expressed as follows: A matroid M is a pair (E,$%
{\cal I}$), where E is a non-empty finite set, and ${\cal I}$ is a non-empty
collection of subsets of E (called independent sets) satisfying the
following properties:

(${\cal I}$ {\it i) }any subset of an independent set is independent;

(${\cal I}${\it \ ii) }if I and J are independent sets with I$\subseteq $ J,
then there is an element $e$ contained in J but not in I, such that I$\cup
\{e\}$ is independent.

A base is defined to be any maximal independent set. By repeatedly using the
property (${\cal I}$ {\it ii) }is straightforward to show that any two bases
have the same number of elements. A subset of E is said to be dependent if
it is not independent. A minimal dependent set is called a circuit. Contrary
to the bases not all circuits of a matroid have the same number of elements.

An alternative definition of a matroid in terms of bases is as follows:

A matroid M is a pair (E, ${\cal B}$), where E is a non-empty finite set and 
${\cal B}$ is a non-empty collection of subsets of E (called bases)
satisfying the following properties:

(${\cal B}$ {\it i) }no base properly contains another base;

(${\cal B}$ {\it ii)} if B$_1$ and B$_2$ are bases and if $h$ is any element
of B$_1,$ then there is an element $g$ of B$_2$ with the property that (B$_1$%
-\{$h$\})$\cup $\{$g$\} is also a base.

It is worth point out that if E is finite set of vectors in a vector space
V, then we can define a matroid on E by taking as bases all linearly
independent subsets of E which span the same subspace as E; a matroid
obtained in this way is called vector matroid.

A matroid can also be defined in terms of circuits:

A matroid M is a pair (E, ${\cal C}$), where E is a non-empty finite set,
and ${\cal C}$ is a collection of a non-empty subsets of E (called circuits)
satisfying the following properties.

(${\cal C}$ {\it i}) no circuit properly contains another circuit;

(${\cal C}$ {\it ii) }if ${\cal C}_1$ and ${\cal C}_2$ are two distinct
circuits each containing an element $c$, then there exists a circuit in $%
{\cal C}_1$ $\cup $ ${\cal C}_2$ which does not contain $c$.

If we start with any of the three definitions (${\cal I}${\it ), }(${\cal B}$%
) and (${\cal C}$) the other two follow as theorems. For instance, it is
possible to prove that (${\cal I}${\it )} implies (${\cal B}$) and (${\cal C}
$). In other words, these three definitions are equivalent. There are other
definitions of a matroid also equivalent to these three, but for the purpose
of this work it is not necessary to consider all of them.

Notice that even from the initial structure of a matroid theory we find
relations such as independent-dependent and base-circuit which suggest
duality. The dual of M, denoted by M$^{*},$ is defined as a pair (E, ${\cal B%
}^{*}$), where ${\cal B}^{*}$ is a non-empty collection of subsets of E
formed with the complements of the bases of M. An immediate consequence of
this definition is that every matroid has a dual and this dual is a unique
matroid. It also follows that the double-dual M$^{**}$ is equal to M.
Moreover, if A is a subset of E, then the size of the largest independent
set contained in A is called the rank of A and is denoted by $\rho $(A). If
M = M$_1$ + M$_2$ and $\rho $(M) = $\rho $(M$_1$) +$\rho $(M$_2$) we shall
say that M is separable. Any maximal non-separable part of M is a component
of M. An important theorem due to Whitney [16] is that if $M_1,..,M_p$ and $%
M_1^{^{\prime }},..,M_p^{^{\prime }}$ are the components of the matroids M
and M' respectively, and if $M_i^{^{\prime }}$ is a dual of $M_i$ (i =
1,...,p), then M' is a dual of M. Conversely, let M and M' be dual matroids,
and let $M_1,..,M_p$ be components of M. Let $M_1^{^{\prime
}},..,M_p^{^{\prime }}$ be the corresponding submatroids of M'. Then $%
M_1^{^{\prime }},..,M_p^{^{\prime }}$ are the components of M', and $%
M_i^{^{\prime }}$ is a dual of $M_i.$

Among the most important matroids we find the binary and regular matroids. A
matroid is binary if it is representable over the integers modulo two. Let
us clarify this definition. An important problem in matroid theory is to see
which matroids can be mapped into some set of vectors in a vector space over
a given field. When such a map exists we speak of a coordinatization (or
representation) of the matroid over the field. This is equivalent to
represent a matroid by a matrix over a given field. (An example of a matroid
that cannot be represented as a matrix is a matroid of rank 3, which has 9
elements \{1,2,3,4,5,6,7,8,9\}and the following 20 circuits: \{7,1,2\},
\{8,1,4\}, \{9,2,3\}, \{7,3,4\}, \{8,3,6\}, \{9,4,5\}, \{7,5,6\}, \{8,2,5\};
\{1,6\}, \{1,9\}, \{6,9\}, \{1,3\}, \{1,5\}, \{2,4\}, \{2,6\}, \{3,5\},
\{4,6\}, \{7,8\}, \{7,9\}, \{8,9\}.) Let GF(q) denote a finite field of
order q. Thus, we can express the definition of a binary matroid as follows:
A matroid which has a coordinatization over GF(2) is called binary.
Furthermore, a matroid which has a coordinatization over every field is
called regular. It turns out that regular matroids play an important role in
matroid theory, among other things, because they play a similar role as
planar graphs do in graph theory [27]. It is known that a graph is planar if
and only if it contains no subgraph homeomorphic to K$_5$ or K$_{3,3}$.
(Recall that K$_n$ is a simple graph in which every pair of distinct
vertices are adjacent, while K$_{r,s},$ where r and s are the number of
vertices in two disjoint sets V$_1$ and V$_2,$ is a complete bipartite graph
in which every vertex of V$_{1}$is joined to every vertex of V$_2$.)
The analogue of this theorem for matroids was provided by Tutte [18]. In
effect, Tutte showed that a matroid is regular if and only if it is binary
and it includes no Fano matroid or its dual. In order to understand this
theorem it is necessary to define the Fano matroid which is some times
referred to as PG(2,2), the projective plane over FG(2). We shall see that
the dual of the Fano matroid is linked with octonions which, at the same
time, are connected to the Englert's compactification of D = 11 supergravity.

A Fano matroid F is the matroid defined on the set E = \{1,2,3,4,5,6,7\}
whose bases are all those subsets of E with three elements except f$_1=$%
\{1,2,4\}, f$_2=$\{2,3,5\}, f$_3=$\{3,4,6\}, f$_4=$\{4,5,7\}, f$_5=$%
\{5,6,1\}, f$_6=$\{6,7,2\} and f$_7=$\{7,1,3\}. The circuits of the Fano
matroid are precisely these subsets and their complements. It follows that
these circuits define the dual F$^{*}$ of the Fano matroid.

Let us write the set E in the form{\it \ }${\cal E}$=$%
\{e_1,e_2,e_{3,}e_{4,}e_5,e_6,e_7\}$. Thus, the subsets used to define the
Fano matroid now become ${\it f}_1=\{e_1,e_2,e_4\}$, ${\it f}%
_2=\{e_2,e_3,e_5\}$, ${\it f}_3=\{e_3,e_4,e_6\}$, ${\it f}_4=\{e_4,e_5,e_7\}$%
, ${\it f}_5=\{e_5,e_6,e_1\}$, ${\it f}_6=\{e_6,e_7,e_2\}$ and ${\it f}%
_7=\{e_7,e_1,e_3\}$. The central idea is to identify the quantities $e_i,$
where $i=1,2,3,4,5,6$ and $7,$ with the octonionic imaginary units.
Specifically, we write an octonion $q$ in the form $%
q=q_0e_0+q_1e_1+q_2e_2+q_3e_3+q_4e_4+q_5e_5+q_6e_6+q_7e_7$. Here, $e_0$
denotes the identity. The product of any two octonions is determined by the
formula 
\begin{equation}
e_ie_j=-\delta _{ij}+\psi _{ijk}e_k  \label{ec.1}
\end{equation}
Here, $\delta _{ij}$ is the Kronecker delta and $\psi _{ijk}$ are fully
antisymmetric structure constants. By taking the tensor $\psi _{ijk}$ equals
1 for each one of the seven combinations {\it f}$_i$ we get all the values
of $\psi _{ijk}.$

The octonion (Cayley) algebra is not associative, but it is alternative.
This means that the basic associator of any three imaginary units is

\begin{equation}
(e_i,e_j,e_k)=(e_ie_j)e_k-e_i(e_je_k)=\varphi _{ijkm}e_m,  \label{ec.2}
\end{equation}
where $\varphi _{ijkl}$ is a fully antisymmetric tensor. It turns out that $%
\varphi _{ijkl}$ and $\psi _{ijk}$ are related by the expression

\begin{equation}
\varphi _{ijkl}=(1/3!)\epsilon _{ijklmnr}\psi _{mnr},  \label{ec.3}
\end{equation}
where $\epsilon _{ijklmnr}$ is the completely antisymmetric Levi-Civita
tensor. It is interesting to note that associating the numerical values
(elements) of the subsets f$_i$ to the indices of $\psi _{mnr}$ and using (%
\ref{ec.3}) we get the other seven subsets of E (with four elements) of the
dual Fano matroid F$^{*}.$ For instance, if we take f$_1,$ then we have $%
\psi _{124}$ and (\ref{ec.3}) gives $\varphi _{3567}$ which leads to the
circuit subset $\{3,5,6,7\}$.

Now, we shall relate the above mathematical structure to the Englert's
octonionic solution [17] of eleven dimensional supergravity. First, let us
introduce the metric

\begin{equation}
g_{ab}=\delta _{ij}h_a^ih_b^j,  \label{ec.4}
\end{equation}
where $h_a^i$ = $h_a^i(x^\mu )$ is a sieben-bein. Here, $x^\mu $ are
coordinates in a patch of the geometrical seven sphere S$^7$. The quantities 
$\psi _{ijk}$ can now be related to the S$^7$ torsion in the form

\begin{equation}
T_{abc}=R_0^{-1}\psi _{ijk}h_a^ih_b^jh_c^k,  \label{ec.5}
\end{equation}
where $R_0$ is the S$^7$ radius. The quantities $\varphi _{ijkl}$ can be
identified with the four-indexed gauge field strength $F_{abcd}$ through the
formula

\begin{equation}
F_{abcd}=R_0^{-1}\varphi _{ijkl}h_a^ih_b^jh_c^kh_d^l.  \label{ec.6}
\end{equation}
Moreover, it is possible to prove that Englert's 7-dimensional covariant
equations can be solved with the identification $F_{abcd}=\lambda
T_{[abc},_{d]},$ where $\lambda $ is a constant. Therefore, $\lambda
T_{abc}=A_{abc}$ is the fully antisymmetric gauge field which is a
fundamental object in 2-brane theory [6].

It is important to mention that in Englert's solution to D = 11 supergravity
the torsion $T_{abc}$ satisfies the Cartan-Schouten equations

\begin{equation}
T_{acd}T_{bcd}=6R_0^{-2}g_{ab},  \label{ec.7}
\end{equation}

\begin{equation}
T_{ead}T_{dbf}T_{fce}=3R_0^{-2}T_{abc.}  \label{ec.8}
\end{equation}
But, as Gursey and Tze [30] noted, these equations are mere septad-dressed,
i.e. covariant forms of the algebraic identities

\begin{equation}
\psi _{ikl}\psi _{jkl}=6\delta _{ij},  \label{ec.9}
\end{equation}

\begin{equation}
\psi _{lim}\psi _{mjn}\psi _{nkl}=3\psi _{ijk}  \label{ec.10}
\end{equation}
respectively. It is worth it to mention that Englert's solution realize the
riemannian curvature-less but torsion-full Cartan geometries of absolute
parallelism on S$^7$.

Let us conclude by making some final comments. In this work, we have shown
that the dual of the Fano matroid is closely related to octonions which at
the same time are essential part of the Englert's solution of absolute
parallelism on S$^7$ of D = 11 supergravity. The Fano matroid and its dual
are the only minimal binary irregular matroids. We know from Hurwitz theorem
(see reference [19]) that octonions is one of the alternative division
algebras (the others are the reals, complex numbers and quaternions). While
among the only parallelizable spheres we find S$^7$ (the others are the
spheres S$^1$ and S$^3$ [31]). This distinctive and fundamental role played
by the Fano matroid, octonions and S$^7$ in such a different areas in
mathematics as combinatorial geometry, algebra and topology respectively
lead us to believe that the relation between these three concepts must have
a deep significate in nature. Of course, it is known that the
parallelizability of S$^1$, S$^3$ and S$^7$ has to do with the existence of
the complex numbers, quaternions and octonions respectively (see reference
[32]). It is also known that using an algebraic topology called K-theory
[33] we find that the only dimensions for division algebras structures on
Euclidean spaces are 1, 2, 4, and 8. We may now add to these remarkable
results another fundamental concept in matroid theory; the Fano matroid. But
besides the importance of the Fano matroid in D = 11 supergravity the
matroid theory offer us the possibility to provide the basis for a duality
principle in M-theory. This is because among other reasons every matroid has
its unique dual matroid. It is interesting to mention that in matroid theory
there is a duality principle [34], which establishes that if A is a
statement in the theory of matroids that has been proved true, then also its
dual A$^{*}$ is true. Perhaps a duality principles such as ``everything in
the physical world is dual for an observer'' or ``the fundamental laws of
physics must be dual'' may constitute the fundamental principles in M-theory.

For further research, it will be interesting to find the exact relation
between D = 11 supergravity and the Fano matroid. It may be also interesting
to see if local supersymmetry is connected with matroids and if matroid
theory may be helpful to find other solutions of D = 11 supergravity [35].
Moreover, it may be of interest to find the connection between
M(atrix)-theory and M(atroid)-theory. At present, we are working in these
problems and we hope to report our results elsewhere.

\medskip\ 

{\bf ACKNOWLEDGMENTS:}

\medskip 

I would like to thank to H. Villegas and N. Alejo for helpful discussions. I
also like to thank the referee for his comments and suggestions. This work
was partially supported by CONACyT grant No. 3898P-E9608.

\smallskip\


\begin{thebibliography}{99}
\bibitem{dual35}  These two problems were suggested by the referee.
\end{thebibliography}
\end{document}